\documentclass[aps,preprint]{revtex4}%
\usepackage{amsfonts}
\usepackage{amsmath}
\usepackage{amssymb}
\usepackage{graphicx}%
\begin{document}
\title{Eavesdropping on the two-way quantum communication protocols with invisible photons}
\author{Qing-yu Cai}
\affiliation{State Key Laboratory of Magnetics Resonance and Atomic and Molecular Physics,
Wuhan Institution of Physics and Mathematics, the Chinese Academy of Sciences,
Wuhan 430071, China}

\begin{abstract}
The crucial issue of quantum communication protocol is its security. In this
paper, we show that all the deterministic and direct two-way quantum
communication protocols, sometimes called ping-pong (PP) protocols, are
insecure when an eavesdropper uses the invisible photon to eavesdrop on the
communication. With our invisible photon eavesdropping (IPE) scheme, the
eavesdropper can obtain full information of the communication with zero risk
of being detected. We show that this IPE scheme can be implemented
experimentally with current technology. Finally, a possible improvement of PP
communication protocols security is proposed.

\end{abstract}

\maketitle

\section{\bigskip introduction}

Quantum communication, or more precisely called quantum key distribution
(QKD), is a physically secure method for the distribution of a secure key
between to distant partners, Alice and Bob, who share a quantum channel and
public authentificated channel [1]. Since the pioneer work of QKD by Bennett
and Brassward in 1984 (BB84) [2], many theoretical proposals [3] and
experimental realizations [4] have been presented. With the efforts of
physicists and engineers, QKD may be the first commercial application of
quantum physics at single photon level [1].

The crucial issue of a QKD protocol is its security. In [5], Brassard,
L\"{u}tkenhaus, Mor and Sanders presented a criteria for a protocol which can
be considered as practical and secure: In order to be practical and secure, a
quantum key distribution scheme must be based on existing---or nearly
existing---technology, but its security must be guaranteed against an
eavesdropper with unlimited computing power whose technology is limited only
by the laws of quantum mechanics. Along this line, we can present an
eavesdropping scheme on the deterministic and direct two-way quantum
communication protocols to gain full information of the key without being
detected. The first deterministic and direct two-way quantum communication
protocol, sometimes called ping-pong (PP) protocol, was presented by
Bostr\"{o}m and Felbinger [6] and its security was extensively studied [7].
After that this PP protocol was enhanced with dense coding feature [8,9]. And
then it was extended to single photon implementation [10-12].

In this paper, an invisible photon eavesdropping (IPE) scheme is presented,
which reveals that the PP-type protocols are insecure under the
eavesdropper-Eve's invisible photon eavesdropping attack. In the IPE scheme,
Eve first selects a photon which is insensitive for Alice and Bob's single
photon detectors, i.e., Eve's such photon is invisible for Alice and Bob. She
adds this invisible photon to the travel pulse and forwards them to Alice.
After Alice's encoding operation, Eve captures her invisible photon and
performs a measurement to draw Alice's encoding message. Because Alice's
single photon detector is insensitive to Eve's invisible photon, Alice and Bob
can not detect Eve's eavesdropping attack when they check the fidelity of
their photons (or the entangled photon pairs). In this way, Eve can gain full
information of the communication without being detected. We show the
experimental possibility of this IPE scheme. Finally, we present a possible
improvement of the PP-type protocols.

\section{Eavesdropping on the PP-type protocols}

Let us start with the brief description of the PP protocol of Bostr\"{o}m and
Felbinger [6]. Bob prepares two photons in an entangled state $|\Psi
^{+}\rangle=(|0\rangle|1\rangle+|1\rangle|0\rangle)/\sqrt{2}$ of the
polarization degree freedom. He keeps one photon (home photon) in his
laboratory, and sends the other photon (travel photon) to Alice through
quantum channel. After receiving the travel photon, Alice randomly switches
the control mode and message mode. In control mode, Alice measures the
polarization photon and announces the measurement result in the classical
public channel. Bob also switches to control mode and measures the home photon
in the same basis after he received Alice's announcements. In the absence of
Eve, both results should be anticorrelated. In the message mode, Alice
performs a unitary operation $Z^{j}$ to encode her message $j\in\{0,1\}$ on
the travel photon, where $Z=|0\rangle\langle0|-|1\rangle\langle1|$. Then Alice
sends the travel photon to Bob. Bob performs a decoding measurement to draw
the information Alice encoded. When the measurement result is $|\Psi
^{+}\rangle$, Bob knows that $j=0$. Likewise, Bob's measurement result
$|\Psi^{-}\rangle$ denotes $j=1$. In order to detect Eve's eavesdropping
hidden in the channel losses and Eve's attack without eavesdropping [7], Alice
and Bob may publish some of their bits to estimate the quantum bit error rate
(QBER) at the end of the key distribution. If both the fidelity of the EPR
pairs and the QBER are normal, Alice and Bob will use their encoding bits to
generate the raw key. After the error-correcting and privacy amplification,
Alice and Bob can get the final key.

However, such final key is not secure because of the potential IPE attack.
Eve's IPE scheme can be described as follow: Eve first prepares a photon in
the states $(|0\rangle+|1\rangle)/\sqrt{2}$. She adds this photon to Bob's
travel photon pulse and forwards them to Alice in line B to A. Since the
number of photons and the polarization of photon are communicative, Eve's
additional photon does not disturb the state of the travel photon. In message
mode, Alice sends the travel photon back to Bob after her encoding operation.
In line A to B, Eve captures her photon and performs a measurement to draw
Alice's encoding message. Eve's measurement result $(|0\rangle+|1\rangle
)/\sqrt{2}$ denotes that Alice's encoded message is $j=0$, and $(|0\rangle
-|1\rangle)/\sqrt{2}$ denotes $j=1$. Sometimes, after receiving the travel
qubit, Alice may switch to control mode. In control mode, Alice measures the
travel photon in the polarization basis. Since Alice's single photon detector
is insensitive to Eve's spy photon, Alice can not find out Eve's photon in
control mode. In this way, Eve can obtain full of Alice's information with
zero risk.

Likewise, Eve can use her IPE scheme in the single photon PP protocol [10-12].
In a single photon PP protocol, Bob first prepares a polarization photon and
then sends it to Alice. After receiving the photon, Alice may switch between
control mode and message mode randomly. In message mode, she encodes her
message on the photon and then sends it back to Bob. Bob performs a
measurement on the photon to gain Alice's encoding message. In control mode,
Alice measures the photon and publishes her measurement results. Bob also
switches to control mode and publishes the state of the photon to estimate the
fidelity of the travel photons. At the end of the key distribution, Alice and
Bob publish some of their encoding and decoding message to estimate the QBER.
In order to gain Alice's encoding information, Eve can add an invisible photon
to the travel photon in line B to A and captures her photon in line A to B.
She measures her photon to gain Alice's encoding message precisely. In this
way, Eve can also obtain full of Alice's information without being detected.

In the case of dense coding protocols [8,9], Eve can also obtain full of
Alice's information with zero risk. In the dense coding protocol, one travel
photon can be encoded with two-bit information. On this occasion, Eve can
prepare an entangled photon pair first. She stores one photon and sends the
other one to Alice to gain information. Also, photons Eve uses are invisible
for Alice.

\section{Experimental possibility of the IPE scheme.}

In experiment, the single photon detector is only sensitive to a special
wavelength, and it can be used near such wavelength. E.g., a commercial
silicon-based single photon detector is used for near infrared wavelengths 600
nm to 900 nm. And a single-electron transistor consisting of a semiconductor
quantum dot in high magnetic field can help one to detect the single
far-infrared photons in the wavelength range 175-210 microm [13]. Thus, Eve
can select a wavelength which is far away from the wavelength Alice and Bob
use. In control mode, Alice can not detect Eve's invisible photon. In message
mode, Alice only performs encoding operations which can be implemented by
using some linear optical elements. Because the number of photons and the
polarization of the photon are communicative, Eve's eavesdropping action does
not disturb the state of Alice and Bob's travel photon. Eve can distinguish
her photon from the travel photon because of the different wavelengths, which
can be realized with a spectroscope experimentally. In this case, Eve can gain
full of Alice's information with zero risk of being detected. In fact, Eve
does not need to add her invisible to Bob's photon pulse. She can send her
invisible photon freely in line B to A and measure her photon in line A to B
after the spectroscope to gain Alice's encoding information. This can be
implemented experimentally with current technology.

\section{Possible modification of the PP-type protocols with filter}

Essentially, such IPE attack is trojan horse attack [14] which aims the
special property of the PP-type quantum communication protocols. The favor of
the PP protocols is that information can be transmitted in a deterministic and
direct way. In order to defeat Eve's IPE attack, the PP-type protocols should
be modified. A possible modification is proposed as below.

Alice adds a filter in her laboratory first [15]. All Bob's photon pulses
should pass through her filter first. Only wavelengths close to the operating
wavelength can be let in. Thus, Eve's invisible photons can be filtered out by
using the filter. If Eve's spy photons can not be filtered out, they will be
detected in Alice's photon detection apparatus. These days, the bandwidth of
optical devices is as narrows as 0.1 to 0.01nm which is comparable to the
laser linewidth. An optical grating to filter out unwanted frequencies may be
used in combination with such the narrow bandwidth devices [15]. In this way,
Alice and Bob can defeat Eve's IPE attack.

\section{remark.}

In [16], Kye, Kim, Kim, and Park (KKKP) have present a QKD protocol with blind
polarization bases. (a.1) Alice chooses a random value of angle $\theta$ and
prepares a photon state with polarization of that angle. She then sends the
qubit to Bob. (a.2) Bob also chooses another random value of angle $\phi$ and
further rotates the polarization direction of the received state by $\phi$ and
then returns to Alice. (a.3) Alice rotates the polarization angle by $-\theta
$, and then encodes the message by rotating the polarization angle $\pm
\frac{\pi}{4}$. And then she sends the photon to Bob. (a.4) After rotating the
polarization angle by $-\phi$, Bob measures the photon to draw Alice's
encoding message. In this way, information can be transmitted directly with
blind bases. In fact, KKKP's protocol is not a two-way ping-pong protocol but
a 3-way protocol, i.e., a key has to travel 3 times between Alice and Bob.
Also, in KKKP's protocol, Eve can first send an n-photon pulse before Alice's
encoding operations, and then performs measurement on her n photons after
receiving her n photons from Alice. Alice can estimate the state of the n
photons precisely by increasing n. However, if Alice's rotating angle $\theta$
is random, Eve can not distinguish which encoding operation Alice performed.
In KKKP, a simple IPE scheme is useless because of the blind bases. Even if
Eve uses the sophisticated photon-number-splitting attack scheme [17,15], she
can only get benefits from the imperfect single-photon sources.

\section{\bigskip summary}

In summary, we have presented an IPE scheme on the PP type quantum
communication protocols. Using our IPE scheme, an eavesdropper can gain full
information of the communication with zero risk. We have shown the
experimental possibility of the IPE scheme. We also suggest the possible
improvements of the PP type protocols to fulfill the condition of the security
against the IPE attack.

\section{acknowledgement}

I am grateful to M. S. Kim for the detailed discussions on this work,
especially the idea of filtering the invisible photons. This work is supported
by the Nature Science Foundation of China under Grant No. 10447140 and 10504039.

\section{references}

[1] N. Gisin, G. Ribordy, W. Tittel, and H. Zbinden, Rev. Mod. Phys. 74, 145 (2002).

[2] C. H. Bennett and G. Brassward, In Procedings IEEE International
Conference on Computers, Systems and Signal Processing , Bangalore, India
(IEEE, New York, 1984), 99. 175-179.

[3] A. K. Ekert, Phys. Rev. Lett. 67, 661 (1991); M. Curty, M. Lewenstein, and
N. L\"{u}kenhaus, Phys. Rev. Lett. 92, 217903 (2004); H.-K. Lo and H. F. Chau,
Science 283, 2050 (1999); D. Mayer, J. Assoc. Comput. Mach. 48, 351 (2001); P.
W. Shor and J. Preskill, Phys. Rev. Lett. 85, 441 (2000).

[4] C. H. Bennett, F. Bessette, G. Brassward, L. Salvail, and J. Smolin, J.
Cryptography 5, 3-28 (1992); For a review, see Ref. [1].

[5] G. Brassward, N. L\"{u}kenhaus, T. Mor, and B. C. Sander, Phys. Rev. Lett.
85, 1330 (2000).

[6] K. Bostr\"{o}m and T. Felbinger, Phys. Rev. Lett. 89, 187902 (2002).

[7] A. W\'{o}jcik, Phys. Rev. Lett. 90, 157901 (2003); Q.-Y. Cai, Phys. Rev.
Lett. 91, 109801 (2003).

[8] Q.-Y. Cai and B.-W. Li, Phys. Rev. A 69, 054301 (2004); Nguyen Ba An,
Phys. Lett. A 328, 6 (2004).

[9] I. P. Degiovanni $et$ $al$., Phys. Rev. A 69, 032310 (2004); A.
W\'{o}jcik, Phys. Rev. A 71, 016301 (2005).

[10] Q.-Y. Cai and B.-W. Li, Chin. Phys. Lett. 21, 601 (2004); H. L\"{u},
X.-D. Yan and X.-Z. Zhang, Chin. Phys. Lett. 21, 2340-2342 (2004).

[11] M. Lucamarini and S. Mancini, Phys. Rev. Lett. 94, 140501 (2005).

[12] F.-G. Deng and G. L. Long, Phys. Rev. A 69, 052319 (2004); Holger
Hoffmann, K. Bostr\"{o}m and T. Felbinger, Phys. Rev. A 72, 016301 (2005);
F.-G. Deng and G. L. Long, Phys. Rev. A 70, 012311 (2004).

[13] S. Komiyama, O. Astafiev, V. V. Antonov, T. Kutsuwa, and H. Hirai, Nature
403, 405 (2000).

[14] N. Gisin, S. Fasel, B. Kraus, H. Zbinden and G. Ribordy,\ quant-ph/0507063.

[15] W.-K. Kye and M. S. Kim, quant-ph/0508028; Private communication with M.
S. Kim.

[16] W.-K. Kye, C.-M. Kim, M. S. Kim, and Y.-J. Park, Phys. Rev. Lett. 95,
040501 (2005).

[17] Q.-Y. Cai, Eavesdropping on the \textquotedblleft quantum key
distribution with blind polarization bases\textquotedblright\ using
sophisticated photon-number-splitting attack, (submitted for publication.).

\section{}
\end{document}